\documentclass{appolb}
\usepackage{bm}
\usepackage{graphicx}

\begin{document}
\title{Deconfinement and Hadron Resonance Gas for Heavy Quarks%
\thanks{Workshop on Criticality in QCD and Hadron Resonance Gas, July 29-31, 2020, Wroclaw, Poland}%
}
\author{Peter Petreczky
\address{Physics Deparment, Brookhaven National Laboratory, Upton, NY 11973, USA}
}
\maketitle
\begin{abstract}
I discuss the deconfinement transition in 2+1 flavor QCD in terms of Polyakov loops as well
as the hadron resonance gas for hadrons containing static quarks and charm quarks.
\end{abstract}
\PACS{12.38.Gc,12.38.Mh,25.75.Nq}
  
\section{Introduction}
Heavy quarks and infinitely heavy (static) quarks  play an important role
when discussing deconfinement transition in strongly interacting matter at high
temperatures. The early works on deconfinement considered the free energy of static quark $Q$
as well as the free energy of static quark antiquark ($Q\bar Q$) pair
\cite{Polyakov:1978vu,Kuti:1980gh,McLerran:1981pb} and the lattice calculations of these
quantities was a focus of many works, see Ref. \cite{Petreczky:2004xs} for a historic review.
Deconfinement is closely related to color screening.
The production rate of quarkonia, bound states of a heavy quark and anti-quark, was suggested 
as a probe of deconfinement in heavy ion collisions \cite{Matsui:1986dk}. The basic idea behind
this proposal was that the color screening in the deconfined medium effects the binding
of heavy quarks (see also Ref. \cite{Bazavov:2009us} for a review). 
Recent lattice QCD studies, however, mostly focus on the chiral aspects of 
the transition at high temperature, see e.g. Refs. \cite{Ding:2015ona,Guenther:2020vqg} for  recent reviews. 
In this contribution I will discuss the deconfinement in 2+1 flavor QCD with (almost) physical
quark masses in terms of Polyakov loops in different renormalization schemes. 

The Hadron Resonance
Gas (HRG) model has been used to understand the thermodynamics below the cross-over temperature
for many years \cite{Karsch:2003vd,Karsch:2003zq,Ejiri:2005wq,Huovinen:2009yb,Huovinen:2009yb,Aoki:2009sc,Bazavov:2017dsy,Huovinen:2017ogf,Fernandez-Ramirez:2018vzu}. The HRG model received significantly
less attention for static and heavy quarks, for example for thermodynamics of charm quarks.
I will discuss the application of HRG model for these
cases, namely the renormalized Polyakov loop and entropy of the static quark and for the charm-baryon number correlations.

\section{The renormalized Polyakov loop and deconfinement}
The expectation value of the Polyakov loop defined as
\begin{equation}
\langle L \rangle = \langle \frac{1}{N} {\rm Tr} {\cal P} \exp(ig \int_0^{1/T} d x_0 A_0(\mathbf{x},x_0) ) \rangle
\end{equation}
is the order parameter for deconfinement in SU(N) gauge theory. It is related to the free energy of a static quark
$L=\exp(-F_Q/T)$ \cite{Polyakov:1978vu,Kuti:1980gh,McLerran:1981pb}. The correlator of the Polyakov loop 
\begin{equation}
C_{PL}(r,T)=\langle L(\mathbf{x}) L^{\dagger}(0) \rangle.
\end{equation}
is related  to the free energy of a static $Q \bar Q$ pair at distance $r=|\mathbf{x}|$ \cite{McLerran:1981pb},
$C_{PL}(r,T)=\exp(-F_{Q\bar Q}(r,T)/T)$.
Color screening in the deconfined phase implies $C_{PL}(r\rightarrow \infty,T)=|\langle L\rangle|^2$ or 
$F_{Q\bar Q}(r\rightarrow \infty,T)=2 F_Q(T)$,
with $F_Q$ being finite.
In the confined (hadronic) phase of SU(N) gauge theory $F_Q$ is infinite and thus $\langle L \rangle=0$. In QCD $\langle L \rangle$ is no longer
an order parameter since $F_Q$ can be finite in the hadronic phase, where it is determined by the binding energy 
of the static-light and the static-strange
hadrons. Despite the fact that the Polyakov loop is not an order parameter in QCD it is still useful for understanding the screening properties
of the medium, as will be discussed below.
In fact, Polyakov loop and its correlator play a key role for the non-perturbative understanding of chromo-electric screening 
\footnote{An alternative approach to study chromo-electric screening in terms of gauge fixed chromo-electric gluon propagators has also been
proposed \cite{Heller:1997nqa,Karsch:1998tx}.}.

\begin{figure}[htb]
\centerline{\includegraphics[width=10cm]{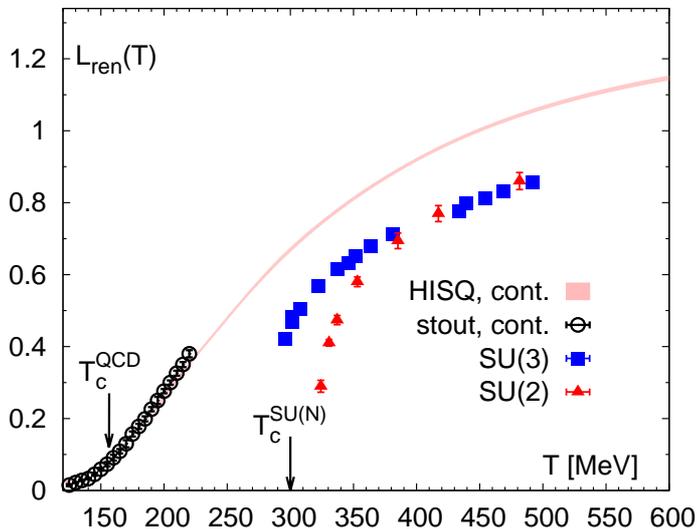}}
\caption{The renormalized Polyakov loop in 2+1 flavor QCD in the continuum limit compared to the Polyakov loop in SU(2) and SU(3)
gauge theories. The arrows indicate the approximate positions of the QCD chiral crossover and the phase transition temperature in SU(2)
and SU(3) gauge theory.}
\label{fig:Lren}
\end{figure}

The free energy of static $Q\bar Q$ pair can be renormalized by requiring that at very short distances it coincides with the zero
temperature $Q\bar Q$ potential \cite{Kaczmarek:2002mc}. This way one also gets the renormalized value of $F_Q$. The renormalized Polyakov loop
in 2+1 f QCD is shown in Fig. \ref{fig:Lren} and compared to the corresponding results in SU(3) \cite{Kaczmarek:2002mc,Kaczmarek:2004gv}
and SU(2) \cite{Digal:2003jc} gauge theories. The calculations in 2+1 flavor QCD have been performed using HISQ action with physical
value of the strange quark mass and light quark masses corresponding to the pion mass of $161$ MeV in the continuum limit \cite{Bazavov:2016uvm}.
Most of the lattice calculations discussed in this contribution have been obtained using this setup.
The continuum extrapolation has been performed \cite{Bazavov:2016uvm} for the Polyakov loop.
I also show the continuum 2+1 flavor results obtained using stout action 
with physical quark masses \cite{Borsanyi:2010bp}.
We see that the behavior of the Polyakov loop in the vicinity of the transition temperature is quite different in QCD and SU(N) gauge theories:
$L_{ren}$ behaves smoothly in QCD and is quite small. An interesting feature of the renormalized Polyakov loop is the fact that it is larger than
one at high temperatures (see Fig. \ref{fig:Lren}) and approaches one from above. This feature can be easily understood if we recall that high
temperature also corresponds to very short  distances, where the zero temperature $Q\bar Q$ potential is given by leading order perturbative
result: $-C_F \alpha_s/r$, and thus is negative. This means that $F_{Q \bar Q}(r,T)$ is also negative at short distances. Because of color 
screening $F_{Q \bar Q}(r,T)$ cannot increase indefinitely with increasing $r$, implying that at high  enough temperature it will be 
negative for all distances, c.f. Fig. 1 (left) in Ref. \cite{Bazavov:2018wmo}. This also means that $F_Q<0$, i.e. $L_{ren}>1$.

\begin{figure}[htb]
\centerline{\includegraphics[width=10cm]{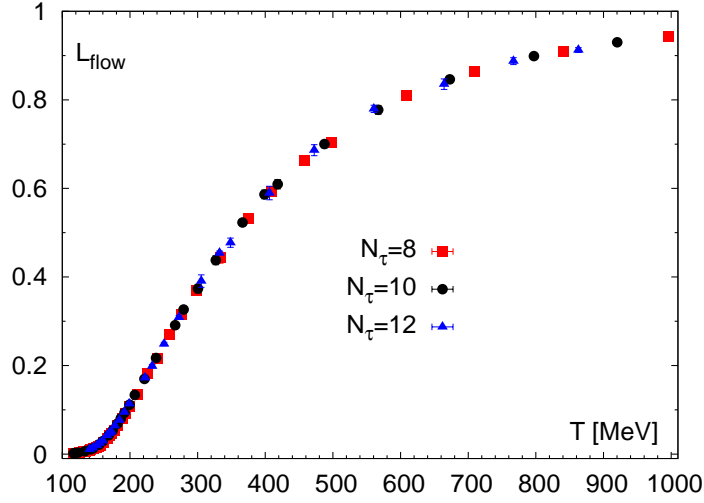}}
\caption{The Polyakov loop calculated with gradient flow for flow time $f=f_0$ at different $N_{\tau}$.}
\label{fig:Lflow}
\end{figure}

Gradient flow provides an alternative way to renormalize the Polyakov loop 
\cite{Petreczky:2015yta,Petreczky:2015qmi,Bazavov:2016uvm}.
The gradient flow is an evolution of the original gauge fields defined on the 4D lattices in a fictitious time, called the flow time, according
to 5D classical equation of motion \cite{Luscher:2010iy}.
This evolution smears the gauge fields in a radius $f=\sqrt{8t}$ and thus removes the UV component
of the fields. It can be thought as continuous smearing of the gauge fields. In order to avoid distortion of thermal physics and to obtain
renormalized results the flow time should satisfy the condition $a \ll f \ll 1/T$ \cite{Petreczky:2015yta}. 
In Fig. \ref{fig:Lflow} I show the renormalized Polyakov loop obtained using the gradient flow for flow time $f=f_0=0.2129$ fm,
and for different
temporal extent $N_{\tau}=1/(aT)$. The fact that there is no significant $N_{\tau}$ dependence in the figure means that the cutoff dependence
is removed. In the crossover region the flow-time dependence of the Polyakov loop is very mild for $f \ge f_0$.
Also notice that the Polyakov loop defined using the gradient flow is smaller than one. This is expected as the renormalized 
Polyakov loop in this case is given by the trace of an SU(3) matrix constructed from the smeared links. The temperature dependence of
the free energy obtained in this scheme is the same as in the conventional renormalization discussed above, the difference amounts to
a temperature independent additive constant \cite{Petreczky:2015yta,Petreczky:2015qmi,Datta:2015bzm,Bazavov:2016uvm}. 
The gradient flow also reduces
the statistical noise, which allows to calculate the Polyakov loop not only for static quarks but also for static charges in higher
representations of SU(3) group (sextet, octet etc.). The Polyakov loops in higher representations are not order parameters for deconfinement
even for SU(N) gauge theories. Nonetheless, they have similar temperature dependence as the Polyakov loop in the fundamental representation
in QCD and are sensitive to color screening, and thus to deconfinement. In the crossover the flow time dependence of the Polyakov
loop in higher representations is somewhat larger and stable results can 
be only obtained for $f\ge 2 f_0$ \cite{Petreczky:2015qmi}.
The Polyakov loops in higher representations satisfy 
the so-called Casimir scaling for $T>300 $ MeV \cite{Petreczky:2015yta}, meaning the corresponding free energies are proportional to
the Casimir operators of the respective representation. This is expected in perturbation theory. In fact Casimir scaling violations
first could show up at order $\alpha_s^4$ \cite{Berwein:2015ayt}. 

\begin{figure}[htb]
\centerline{
\includegraphics[width=7cm]{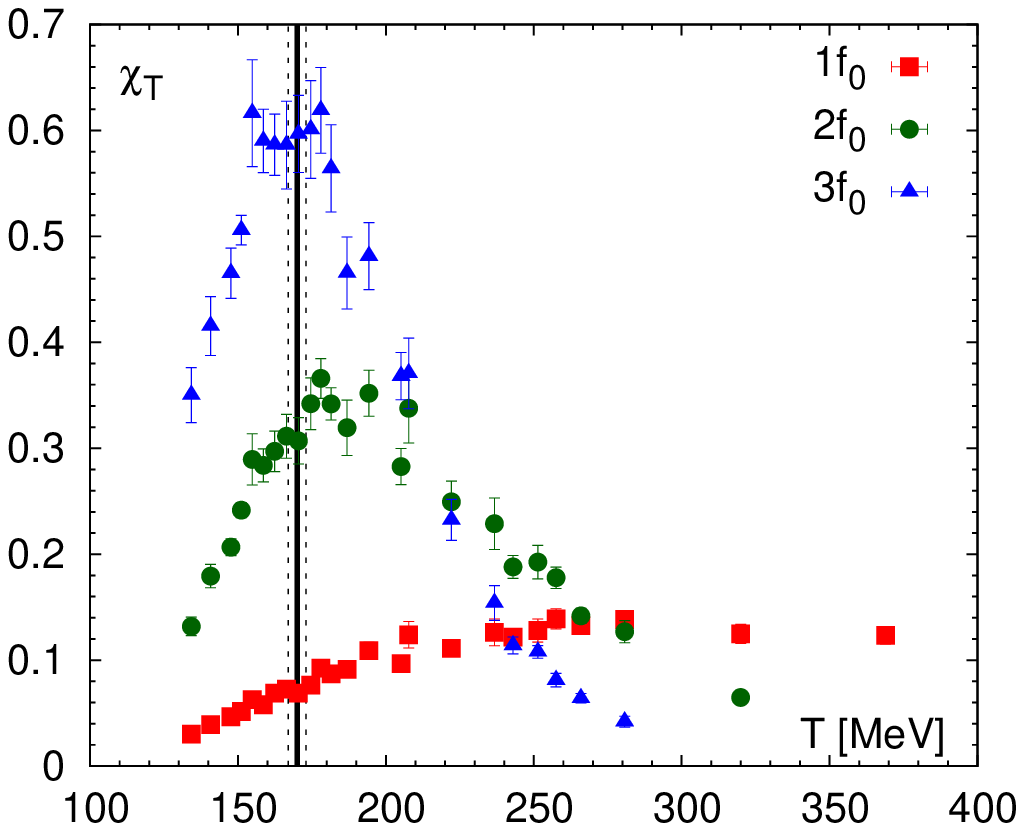}
\includegraphics[width=7cm]{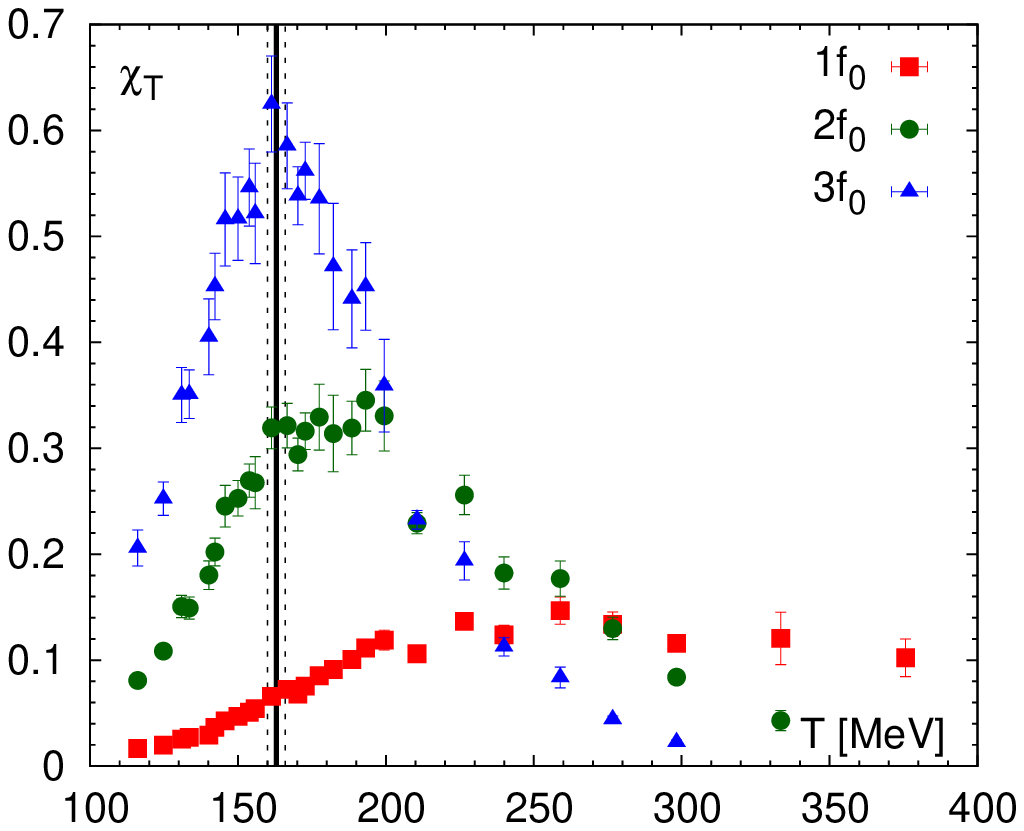}
}
\caption{The Polyakov loop susceptibility $\chi_T$ calculated for flow times $f=f_0,~2 f_0$ and $3f_0$ on $N_{\tau}=6$ lattice (left)
and on $N_{\tau}=8$ lattice (right).}
\label{fig:chiT}
\end{figure}

Since the gradient flow reduces the statistical errors and at the same time renormalizes composite operators it can
be used to study the renormalized Polyakov loop susceptibilities. Following Ref. \cite{Lo:2013etb} 
three types of susceptibilities can be defined:
\begin{eqnarray}
&
\chi=(VT^3) (\langle L^2 \rangle - \langle L \rangle^2 ),\\[3mm]
&
\chi_L=(VT^3) (\langle ({\rm Re} L)^2 \rangle - \langle L \rangle^2),~\chi_T= V T^3 \langle ({\rm Im} L)^2 \rangle.
\end{eqnarray}
Note that in QCD $\langle {\rm Im} L \rangle=0$. The flow dependence of $\chi$ and $\chi_L$ turns out to be significant 
\cite{Bazavov:2016uvm}. For $f=3 f_0$ these quantities show a peak around $T \simeq 180-200$ MeV. In Fig. \ref{fig:chiT}
I show the lattice results for $\chi_T$. This quantity also has a significant flow time dependence. For $f=3f_0$
it has a peak around the chiral cross-over temperature. The flow time dependence is largely reduced in the ratio
$\chi_T/\chi$ \cite{Bazavov:2016uvm}. This ratio has a characteristic decrease around the chiral cross-over temperature \cite{Bazavov:2016uvm}.
In this sense $\chi_T$ may be sensitive to deconfinement in QCD.

\begin{figure}[htb]
\centerline{
\includegraphics[width=10cm]{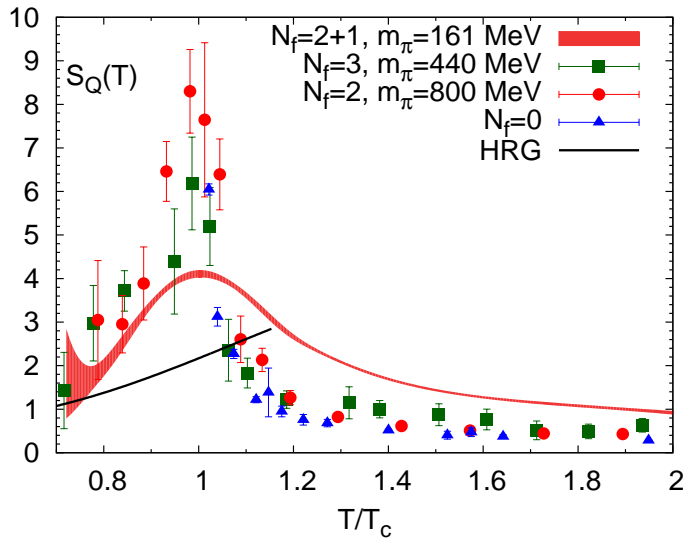}
}
\caption{The continuum estimate for the entropy of a static quark, $S_Q$ in 2+1 flavor QCD (red band) compared
to the corresponding results in 3-flavor and 2-flavor QCD with unphysical quark masses as well as in SU(3)
gauge theory (zero flavor QCD), see text.}
\label{fig:SQ}
\end{figure}

At low temperatures one may try to understand the Polyakov loop in terms of a gas of static-light and static-strange 
hadrons \cite{Megias:2012kb,Bazavov:2013yv}. 
Hadrons consisting of a static quark will interact with other hadrons in the medium and one can assume that these interactions 
can be taken care of by including static-light and static-strange resonances in the spirit of the HRG model \cite{Bazavov:2013yv}. 
Lattice QCD calculations provide information on the few low lying static light hadron states, which is not sufficient. We could use 
charm and beauty hadrons as proxies for static-light and static-strange hadrons once the finite quark effect 
has been taken into account \cite{Bazavov:2013yv}.
However, the knowledge of the spectrum of charm and beauty hadrons is rather incomplete. Therefore, in order to include higher excited
states one needs to use the quark model results \cite{Godfrey:1985xj,Godfrey:1986wj,Ebert:1997nk,Ebert:2011zz}
for the hadron spectrum \cite{Bazavov:2013yv}. To compare $F_Q$ obtained in such HRG model with the 
continuum extrapolated lattice results one needs to shift it by an additive constant \cite{Bazavov:2013yv}. The comparison of
the lattice and the HRG result shows that HRG can only describe the free energy of a static quark for $T<140$ MeV. At higher temperatures
$F_Q$ shows a qualitatively different behavior. In particular it has an inflection point around the chiral cross-over temperature \cite{Bazavov:2013yv}.
Since $F_Q$ is a physical quantity this inflection point could be related to the deconfinement transition. Unlike the inflection point
for $L_{ren}$ it does not depend on the choice of the renormalization scheme. The entropy of a static quark is defined as
\begin{equation}
S_Q=-\frac{\partial F_Q}{\partial T}.
\end{equation}
The inflection point in $F_Q$ corresponds to a maximum in $S_Q$. The comparison of the lattice results and the HRG results is simpler for $S_Q$
because the additive normalization constant drops out. The comparison of the lattice results with HRG for $S_Q$ is shown in Fig. \ref{fig:SQ}.
The temperature axis in the figure has been rescaled by $T_c=156.5$ MeV \cite{Bazavov:2018mes} for the HRG result and by 159.5 MeV for
the 2+1 flavor QCD results, since the lattice calculations have been performed for $m_l=m_s/20$ instead of the physical value $m_l=m_s/27$.
This results in 3 MeV upward shift in $T_c$ according to the analysis of Ref. \cite{Bazavov:2011nk}. The lattice result for $S_Q$ clearly disagrees
with HRG. The entropy of the static quark shows a peak around the chiral cross-over temperature. For comparison we also show lattice results for
$S_Q$ from SU(3) gauge theory \cite{Kaczmarek:2002mc,Kaczmarek:2004gv}, 3-flavor QCD \cite{Petreczky:2004pz} and 
and 2-flavor QCD \cite{Kaczmarek:2005gi} with larger than the physical quark mass. 
The temperature variables in the corresponding lattice results have been rescaled
by the phase transition temperature of SU(3) gauge theory and by the chiral cross-over temperature for 3-flavor and 2 flavor QCD
for the respective quark masses. For SU(3) gauge theory the divergence in $S_Q$ for $T \rightarrow T_c^{+}$ is clearly related
to the deconfinement transition. For 3-flavor and 2-flavor QCD at larger than physical quark masses we know that the chiral
crossover temperature and the deconfinement temperature defined in terms of Polyakov loop susceptibility coincide \cite{Karsch:2000kv}.
So here it is also justified to associate the peak in $S_Q$ with the deconfinement transition. From these considerations we conclude
that the peak in $S_Q$ in 2+1 flavor QCD is also associated with the deconfinement temperature and the chiral and deconfinement
transitions coincide is that sense for the physical value of the quark masses. Another interesting question is whether the critical
behavior in 2+1 flavor QCD for $m_l \rightarrow 0$ has an imprint on the Polyakov loop expectation value. Very recent lattice calculations
suggest that this is indeed the case \cite{Clarke:2020htu}. Since the chiral phase transition temperature $T_c=132(+6)(-3)$ MeV
\cite{Ding:2019prx} is significantly
smaller than the chiral cross-over temperature for physical quark masses it is possible that $F_Q$ as a function of temperature
has two inflection points \footnote{I thank F. Karsch for raising this point during my talk.}.

\section{Charm-baryon number correlations}
Charm quark fluctuations and charm baryon number correlations up to the fourth order have been studied in 2+1 flavor QCD on the lattice
using HISQ action \cite{Bazavov:2014yba}. The charm fluctuations and charm baryon number correlations are defined
through the derivatives of the QCD pressure with respect to the corresponding chemical potentials as
\begin{equation}
\chi_n^C=\frac{\partial^n (p/T^4)}{\partial (\mu_C/T)^n},~
\chi_{nm}^{BC}=\frac{\partial^{n+m}(p/T^4)}{\partial (\mu_C/T)^n \partial(\mu_B/T)^m}.
\end{equation}
Because charm quarks are heavy only $|C|=1$ sector contributes to the above quantities in the temperature
range of interest. The ratio $\chi_{13}^{BC}/\chi_{22}^{BC}$ is sensitive to deconfinement. Below the
crossover this ratio is one because in HRG framework the relevant degrees of freedom are singly charmed baryons, $|B|=|C|=1$.
At high temperatures this ratio approaches three, since the relevant degrees of freedom are charm quarks. 
The lattice calculations confirm these expectations \cite{Bazavov:2014yba}. 
Let 
me mention that the above ratio is convenient because the lattice artifacts as well as the uncertainties
related to the tuning of the charm quark mass cancel out. For a detailed comparison with HRG it is also convenient to
consider a ratio for the same reasons. The relevant ratio in this case is $\chi_{13}^{BC}/(\chi_2^C-\chi_{13}^{BC})$,
which can be considered as a proxy for charm baryon to charm meson pressure \cite{Bazavov:2014yba}.
As mentioned before the charmed hadron
spectrum is not very well known. This is especially the case
for charmed baryons. Therefore, if one only includes charm hadrons from Particle Data Group the HRG largely 
under-predicts the lattice results for $\chi_{13}^{BC}/(\chi_2^C-\chi_{13}^{BC})$. If one includes
additional charmed hadrons from quark models, as discussed for the static quark free energy, the lattice results
agree with HRG below and in the vicinity of the chiral crossover \cite{Bazavov:2014yba}. 

Another interesting question is the nature of charm
degrees of freedom above the chiral cross-over temperature but for $T<250$ MeV. In this temperature region
the baryon number charm correlations are not described by HRG but also are significantly smaller than in the 
quark gas \cite{Bazavov:2014yba}. One may wonder if charm hadron like excitations can explain this feature. It is
known that charmonia can exist above the deconfinement phase transition temperature \cite{Asakawa:2003re,Datta:2003ww}.
There are some indication that hadron like excitations may exist in the deconfined phase also in the light quark
sector \cite{Wetzorke:2001dk,Karsch:2002wv,Asakawa:2002xj}. Therefore, in Ref. \cite{Mukherjee:2015mxc}  it
was proposed that the lattice results on baryon number charm correlations can be understood if charm hadron
like excitations exist above $T_c$. According to this model charm quark are dominant degrees of freedom  only
for $T>200$ MeV \cite{Mukherjee:2015mxc}.

\section{Conclusions}

From the discussions above it is clear that the Polyakov loop behaves quite differently in QCD and SU(N) gauge theory.
In particular, the renormalized Polyakov loop in QCD is small at the crossover temperature irrespective of
the renormalization scheme and shows a smooth behavior. Gradient flow can be used to renormalize the Polyakov loop and
study its fluctuations, as well as Polyakov loops in higher representations. The Polyakov loop susceptibilities 
do not show the type of behavior seen in SU(N) gauge theories and therefore cannot be used to define the deconfinement
transition temperature. On the other hand the entropy of a static quark can be used to identify the deconfinement temperature
through its maximum. The entropy of the static quark has a peak around the chiral cross-over temperature. In this sense
the chiral and deconfinement transitions coincide for the physical values of the quark masses. The HRG model fails to describe
the free energy of static quarks for $T>140$ MeV despite the fact that many resonances have been included in the analysis.
This may be due to the deconfinement physics encoded in the Polyakov loop.
The charm-baryon number correlations can be described by the HRG model once additional hadron states, that are not yet observed experimentally 
but predicted by the quark model, are included. In this respect charm and static quarks are quite different.
For $T_c<T<250$ MeV the baryon-charm correlations cannot be described by
HRG model but are very different from the quark gas expectations. This feature may be explained if one assumes charm hadron like
excitations to exist above the cross-over temperature.


\section*{Acknowledgements}
This work has been supported by the U.S Department of Energy 
through grant contract No. DE-SC0012704.

\bibliographystyle{h-physrev.bst}
\bibliography{ref}

\begin{thebibliography}{10}

\bibitem{Polyakov:1978vu}
A.~M. Polyakov,
\newblock Phys. Lett. {\bf B72}, 477 (1978).

\bibitem{Kuti:1980gh}
J.~Kuti, J.~Polonyi, and K.~Szlachanyi,
\newblock Phys. Lett. {\bf B98}, 199 (1981).

\bibitem{McLerran:1981pb}
L.~D. McLerran and B.~Svetitsky,
\newblock Phys. Rev. {\bf D24}, 450 (1981).

\bibitem{Petreczky:2004xs}
P.~Petreczky,
\newblock Nucl. Phys. B Proc. Suppl. {\bf 140}, 78 (2005), hep-lat/0409139.

\bibitem{Matsui:1986dk}
T.~Matsui and H.~Satz,
\newblock Phys. Lett. B {\bf 178}, 416 (1986).

\bibitem{Bazavov:2009us}
A.~Bazavov, P.~Petreczky, and A.~Velytsky,
\newblock {\em {Quarkonium at Finite Temperature}} (, 2010), pp. 61--110,
  0904.1748.

\bibitem{Ding:2015ona}
H.-T. Ding, F.~Karsch, and S.~Mukherjee,
\newblock Int. J. Mod. Phys. {\bf E24}, 1530007 (2015), 1504.05274.

\bibitem{Guenther:2020vqg}
J.~N. Guenther,
\newblock (2020), 2010.15503.

\bibitem{Karsch:2003vd}
F.~Karsch, K.~Redlich, and A.~Tawfik,
\newblock Eur. Phys. J. C {\bf 29}, 549 (2003), hep-ph/0303108.

\bibitem{Karsch:2003zq}
F.~Karsch, K.~Redlich, and A.~Tawfik,
\newblock Phys. Lett. B {\bf 571}, 67 (2003), hep-ph/0306208.

\bibitem{Ejiri:2005wq}
S.~Ejiri, F.~Karsch, and K.~Redlich,
\newblock Phys. Lett. B {\bf 633}, 275 (2006), hep-ph/0509051.

\bibitem{Huovinen:2009yb}
P.~Huovinen and P.~Petreczky,
\newblock Nucl. Phys. A {\bf 837}, 26 (2010), 0912.2541.

\bibitem{Aoki:2009sc}
Y.~Aoki {\em et~al.},
\newblock JHEP {\bf 06}, 088 (2009), 0903.4155.

\bibitem{Bazavov:2017dsy}
A.~Bazavov, P.~Petreczky, and J.~H. Weber,
\newblock Phys. Rev. {\bf D97}, 014510 (2018), 1710.05024.

\bibitem{Huovinen:2017ogf}
P.~Huovinen and P.~Petreczky,
\newblock Phys. Lett. B {\bf 777}, 125 (2018), 1708.00879.

\bibitem{Fernandez-Ramirez:2018vzu}
C.~Fern\'andez-Ram\'\i{}rez, P.~M. Lo, and P.~Petreczky,
\newblock Phys. Rev. C {\bf 98}, 044910 (2018), 1806.02177.

\bibitem{Heller:1997nqa}
U.~M. Heller, F.~Karsch, and J.~Rank,
\newblock Phys. Rev. {\bf D57}, 1438 (1998), hep-lat/9710033.

\bibitem{Karsch:1998tx}
F.~Karsch, M.~Oevers, and P.~Petreczky,
\newblock Phys. Lett. {\bf B442}, 291 (1998), hep-lat/9807035.

\bibitem{Kaczmarek:2002mc}
O.~Kaczmarek, F.~Karsch, P.~Petreczky, and F.~Zantow,
\newblock Phys. Lett. {\bf B543}, 41 (2002), hep-lat/0207002.

\bibitem{Kaczmarek:2004gv}
O.~Kaczmarek, F.~Karsch, F.~Zantow, and P.~Petreczky,
\newblock Phys. Rev. {\bf D70}, 074505 (2004), hep-lat/0406036,
\newblock [Erratum: Phys. Rev.D72,059903(2005)].

\bibitem{Digal:2003jc}
S.~Digal, S.~Fortunato, and P.~Petreczky,
\newblock Phys. Rev. {\bf D68}, 034008 (2003), hep-lat/0304017.

\bibitem{Bazavov:2016uvm}
A.~Bazavov {\em et~al.},
\newblock Phys. Rev. {\bf D93}, 114502 (2016), 1603.06637.

\bibitem{Borsanyi:2010bp}
Wuppertal-Budapest, S.~Borsanyi {\em et~al.},
\newblock JHEP {\bf 09}, 073 (2010), 1005.3508.

\bibitem{Bazavov:2018wmo}
TUMQCD, A.~Bazavov, N.~Brambilla, P.~Petreczky, A.~Vairo, and J.~H. Weber,
\newblock Phys. Rev. D {\bf 98}, 054511 (2018), 1804.10600.

\bibitem{Petreczky:2015yta}
P.~Petreczky and H.~P. Schadler,
\newblock Phys. Rev. {\bf D92}, 094517 (2015), 1509.07874.

\bibitem{Petreczky:2015qmi}
H.-P. Schadler and P.~Petreczky,
\newblock PoS {\bf LATTICE2015}, 163 (2016), 1511.04591.

\bibitem{Luscher:2010iy}
M.~Luscher,
\newblock JHEP {\bf 08}, 071 (2010), 1006.4518,
\newblock [Erratum: JHEP03,092(2014)].

\bibitem{Datta:2015bzm}
S.~Datta, S.~Gupta, and A.~Lytle,
\newblock Phys. Rev. D {\bf 94}, 094502 (2016), 1512.04892.

\bibitem{Berwein:2015ayt}
M.~Berwein, N.~Brambilla, P.~Petreczky, and A.~Vairo,
\newblock Phys. Rev. {\bf D93}, 034010 (2016), 1512.08443.

\bibitem{Lo:2013etb}
P.~M. Lo, B.~Friman, O.~Kaczmarek, K.~Redlich, and C.~Sasaki,
\newblock Phys. Rev. {\bf D88}, 014506 (2013), 1306.5094.

\bibitem{Megias:2012kb}
E.~Megias, E.~Ruiz~Arriola, and L.~L. Salcedo,
\newblock Phys. Rev. Lett. {\bf 109}, 151601 (2012), 1204.2424.

\bibitem{Bazavov:2013yv}
A.~Bazavov and P.~Petreczky,
\newblock Phys.Rev. {\bf D87}, 094505 (2013), 1301.3943.

\bibitem{Godfrey:1985xj}
S.~Godfrey and N.~Isgur,
\newblock Phys.Rev. {\bf D32}, 189 (1985).

\bibitem{Godfrey:1986wj}
S.~Godfrey and R.~Kokoski,
\newblock Phys.Rev. {\bf D43}, 1679 (1991).

\bibitem{Ebert:1997nk}
D.~Ebert, V.~Galkin, and R.~Faustov,
\newblock Phys.Rev. {\bf D57}, 5663 (1998), hep-ph/9712318.

\bibitem{Ebert:2011zz}
D.~Ebert, R.~N. Faustov, and V.~O. Galkin,
\newblock PoS {\bf QCD-TNT-II}, 016 (2011).

\bibitem{Bazavov:2018mes}
HotQCD, A.~Bazavov {\em et~al.},
\newblock Phys. Lett. B {\bf 795}, 15 (2019), 1812.08235.

\bibitem{Bazavov:2011nk}
A.~Bazavov {\em et~al.},
\newblock Phys. Rev. {\bf D85}, 054503 (2012), 1111.1710.

\bibitem{Petreczky:2004pz}
P.~Petreczky and K.~Petrov,
\newblock Phys. Rev. {\bf D70}, 054503 (2004), hep-lat/0405009.

\bibitem{Kaczmarek:2005gi}
O.~Kaczmarek and F.~Zantow,
\newblock (2005), hep-lat/0506019.

\bibitem{Karsch:2000kv}
F.~Karsch, E.~Laermann, and A.~Peikert,
\newblock Nucl. Phys. {\bf B605}, 579 (2001), hep-lat/0012023.

\bibitem{Clarke:2020htu}
D.~A. Clarke, O.~Kaczmarek, F.~Karsch, A.~Lahiri, and M.~Sarkar,
\newblock (2020), 2008.11678.

\bibitem{Ding:2019prx}
H.~Ding {\em et~al.},
\newblock Phys. Rev. Lett. {\bf 123}, 062002 (2019), 1903.04801.

\bibitem{Bazavov:2014yba}
A.~Bazavov {\em et~al.},
\newblock Phys.Lett. {\bf B737}, 210 (2014), 1404.4043.

\bibitem{Asakawa:2003re}
M.~Asakawa and T.~Hatsuda,
\newblock Phys. Rev. Lett. {\bf 92}, 012001 (2004), hep-lat/0308034.

\bibitem{Datta:2003ww}
S.~Datta, F.~Karsch, P.~Petreczky, and I.~Wetzorke,
\newblock Phys. Rev. {\bf D69}, 094507 (2004), hep-lat/0312037.

\bibitem{Wetzorke:2001dk}
I.~Wetzorke, F.~Karsch, E.~Laermann, P.~Petreczky, and S.~Stickan,
\newblock Nucl. Phys. Proc. Suppl. {\bf 106}, 510 (2002), hep-lat/0110132.

\bibitem{Karsch:2002wv}
F.~Karsch {\em et~al.},
\newblock Nucl. Phys. A {\bf 715}, 701 (2003), hep-ph/0209028.

\bibitem{Asakawa:2002xj}
M.~Asakawa, T.~Hatsuda, and Y.~Nakahara,
\newblock Nucl. Phys. B Proc. Suppl. {\bf 119}, 481 (2003), hep-lat/0208059.

\bibitem{Mukherjee:2015mxc}
S.~Mukherjee, P.~Petreczky, and S.~Sharma,
\newblock Phys. Rev. {\bf D93}, 014502 (2016), 1509.08887.

\end{thebibliography}

\end{document}